\documentclass[twocolumn,letter]{jpsj3} 
\usepackage{txfonts}

\title{
Spin-Polarization in Magneto-Optical Conductivity of Dirac Electrons
}

\author{Yuki Fuseya\thanks{E-mail address: fuseya@mp.es.osaka-u.ac.jp} 
 $^{1} $,
 Masao Ogata $^{2}$,
 and 
 Hidetoshi Fukuyama$^{3}$
}
\inst{
$^{1}$ Department of Materials Engineering Science, Osaka University, Toyonaka, Osaka 560-8531\\
$^{2}$ Department of Physics, University of Tokyo, 7-3-1 Hongo, Bunkyo-ku, Tokyo 113-0033\\
$^{3}$ Department of Applied Physics and Research Institute for Science and Technology, Tokyo University of Science, Kagurazaka, Shinjuku-ku, Tokyo 162-860
}

\abst{
	A mechanism is proposed based on the Kubo formula to generate a spin-polarized magneto-optical current of Dirac electrons in solids which have strong spin-orbit interactions such as bismuth.
	The ac current response functions are calculated in the isotropic Wolff model under an external magnetic field, and the selection rules for Dirac electrons are obtained.
	By using the circularly polarized light and tuning its frequency, one can excite electrons concentrated in the spin-polarized lowest Landau level when the chemical potential locates in the band gap, so that spin-polarization in the magneto-optical current can be achieved.
}

\kword{Dirac electron, spin polarized electric current, bismuth, Kubo formula}

\usepackage{amsmath,bm,mathrsfs}
\renewcommand{\Im}{{\rm i}}

\newcommand{\D}{\Delta}

\newcommand{\rc}{{\rm c}}
\newcommand{\rv}{{\rm v}}

\newcommand{\bk}{\bm{k}}
\newcommand{\bp}{\bm{p}}

\newcommand{\bpi}{\bm{\pi}}

\newcommand{\wc}{\omega_{\rm c}^*}

\newcommand{\scr}[1]{\mathscr{#1}}
\newcommand{\ve}{\varepsilon}

\begin{document}
\maketitle
	%
	The motion of electrons can be described by a Dirac type equation when there are two bands coupled by a strong spin-orbit interaction as shown by Wolff in the study of bismuth and its alloys\cite{Wolff1964,Cohen1960}.
	%
	%
	The validity of this approach has been confirmed experimentally\cite{Dresselhaus1971,Edelman1976,Vecchi1976,Zhu2011}.
	%
	%
	%
	%
	In the Wolff Hamiltonian (three-dimensional $4\times 4$ matrix and massive), the spin-orbit interaction is included in the off-diagonal elements as in the original Dirac Hamiltonian in relativistic quantum mechanics.
	These spin-dependent off-diagonal elements lead to non-trivial inter-band effects such as large diamagnetism\cite{Fukuyama1970} and unconventional inter-band Hall effects\cite{Fuseya2009}.
	Although the graphene-type Hamiltonian (e.g., in graphene\cite{Zheng2002,Gusynin2006,Fukuyama2007} or $\alpha$-ET$_2$I$_3$\cite{Kobayashi2008}) also gives Dirac electrons, they are intrinsically different from the Wolff Hamiltonian with respect to the spin properties.
	The graphene Hamiltonian (two-dimensional $2\times 2$ matrix and massless) is not related to the real spin, but to the pseudo spin, which expresses the degrees of freedom of the sublattice.
	%
	%
	It has nothing to do with spin-orbit interactions or spin properties. 
	Wolff discussed in his Hamiltonian that the spin-orbit interaction gives rise to photo-induced spin transitions in magneto-optical measurements. 
	This spin transition itself was an important new feature at that time, but no particular spin phenomena have been proposed so far using the spin-orbit interaction or the spin transition.
	%
\begin{figure}
\begin{center}
\includegraphics[width=8.0cm]{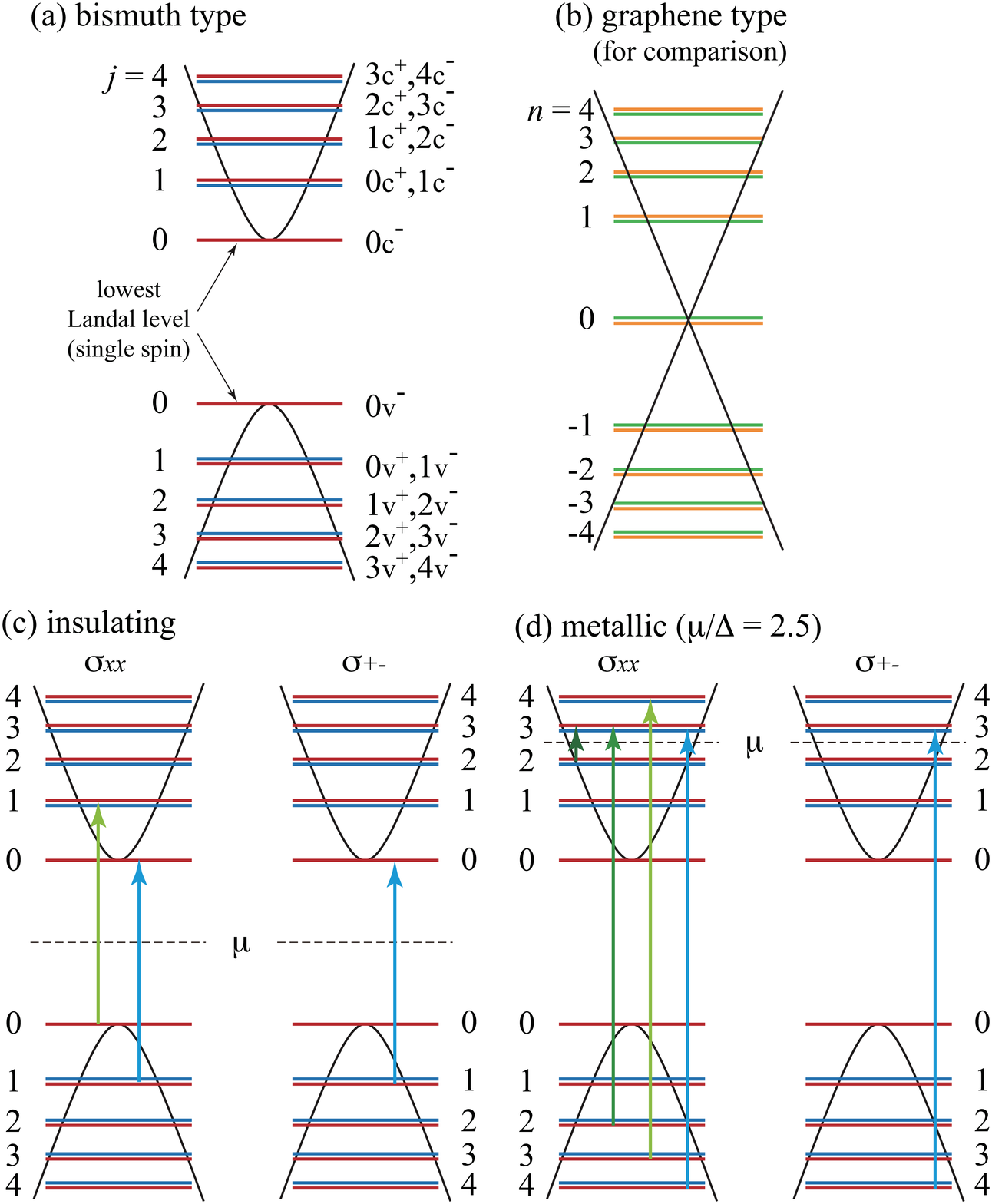}
\end{center}
\caption{(color online) (a) Illustrations of Landau levels for bismuth. The numbers beside the levels denoting the total angular momentum $j$, the $+$ and $-$ correspond to $\sigma$, and ``c" and ``v" are the abbreviations of conduction and valence band, respectively.
(b) (for comparison) The Landau levels for graphene, which are given by $E_n = {\rm sgn} (n) \gamma \sqrt{2eB|n|/c}$, where $\gamma$ is the velocity\cite{Zheng2002}. There, double degeneracy is due to the degrees of freedom of sublattice pseudo spin, not the real spin.
The possible transitions from the selection rule are displayed by arrows for (c) insulating ($|\mu| < \D$), and (d) metallic ($\mu/\D=2.5$) regions.

}
\label{LL}
\end{figure}
	%
	%
	
	In the present Letter, we propose a mechanism to generate a spin-polarized electric current by use of the lowest Landau level and the circularly polarized light. 
	Before showing the details of calculations, we illustrate the phenomena in terms of the Landau levels (Fig. \ref{LL}).
	In the presence of strong spin-orbit interaction as in bismuth, namely, in Wolff Hamiltonian, the energy dispersion when there is no magnetic field ($B=0$), and the Landau levels in the case of $B>0$ are shown in Fig. \ref{LL} (a).
	The Landau levels are given by $E_j = \pm \sqrt{\D^2 + 2\D \wc j}$, where $2\D$ is the band gap and $\wc$ is the renormalized cyclotron frequency (origin of energy is taken at the center of the band gap)\cite{Cohen1960,Wolff1964}.
	%
	%
	The interval of Landau levels decreases as $j$ increases.
	The total angular momentum $j$ is related to the orbital- and spin-quantum numbers as $j=n + 1/2 + \sigma/2$, where $\sigma=+1$ and $-1$ correspond to the up and down spin of electrons.
	For $j \geq1$, the Landau levels with $n$ and $\sigma=-1$, and with $n-1$ and $\sigma=+1$ are degenerate.
	On the other hand, the lowest Landau level with $j=0$ permits only the state with $n=0$ and  $\sigma=-1$.
	This lowest Landau level with a spin of particular direction is a striking characteristic of the Wolff Hamiltonian.
	Such a state cannot be obtained in the case of gapless graphene Hamiltonian as in Fig. \ref{LL} (b).

	We discuss optical conductivities in the Landau levels shown in Fig. \ref{LL} (a).
	%
	%
	Since the dipole matrix elements are relevant to the optical conductivity, it turns out that the transitions of $\D j = \pm 1$ are allowed in the optical conductivity, $\sigma_{xx}(\omega)$, as is depicted in the left panels of Figs. \ref{LL} (c) and (d).
	In the insulating state (Fig. \ref{LL} (c)), we see that there are two processes for the smallest excitation energy involving $j=0$ states.
	In this case we do not have net spin polarization.
	%
	%
	%
	On the other hand, if we use a circularly polarized light, i.e., in the case of $\sigma_{+-}(\omega)$, the allowed excitations are restricted to that with $\D j = -1$ by the selection rule. 
	Therefore, for the smallest excitation energy, only the transition shown in the right panel of Fig. \ref{LL} (c) is allowed, which leads to the spin polarization.
	%
	%
		
	For our spin-polarized electric current, the spin-orbit interaction is essential (actually, the spin transition always accompanies the orbital transition), and the Landau levels are characterized by $j$ not by $n$.
	Our mechanism uses (1) the inter-band transition, (2) the lowest Landau level, and (3) the circularly polarized light, whereas Wolff mainly investigated the intra-band transition and did not consider the specific property of the lowest Landau level.
	In graphene, in contrast, it is not possible to obtain the spin polarization by the present mechanism, since its Hamiltonian does not have spin-orbit interactions or spin transitions.
	Note also that the present spin-polarized current is (i) a current going through the bulk and not along the surface, (ii) accompanied by an electric current (i.e., not a pure spin current), and (iii) realized under the magnetic field.
	These properties are different from the spin Hall effects\cite{Murakami2003,Sinova2004}, the quantum spin Hall effects or the surface state of the topological insulators\cite{Kane2005}.
	%
	%

	
	In the following, we show explicit calculations based on the ``isotropic" Wolff Hamiltonian.
	As an effective model of bismuth, Cohen and Blount employed a simple one electron Hamiltonian with the spin-orbit interaction\cite{Cohen1960}:
	\begin{align}
		\mathscr{H}
		&= \frac{p^2}{2m}+ V + \frac{\nabla^2 V}{8m^2 c^2}+\frac{1}{4m^2c^2} \bm{p} \cdot \left(\bm{\sigma} \times \bm{\nabla}V\right).
		\label{Hamiltonian}
\end{align}
	The Pauli matrix $\bm{\sigma}$ corresponds to the real spin of electrons, $V$ is the crystal potential, and the last term expresses the spin-orbit interaction.
	Wolff found that eq. (\ref{Hamiltonian}) can be written in an essentially identical form to the $4\times4$ Dirac Hamiltonian on the basis of the $\bk \cdot \bp$ theory\cite{Wolff1964}.
	The velocity of bismuth is anisotropic by nature.
	However in this Letter,  we discuss the case with isotropic velocity as
	\begin{align}
		\mathscr{H}
		&=\begin{pmatrix}
		\D & \Im \gamma \bk \cdot \bm{\sigma}\\
		-\Im \gamma \bk \cdot \bm{\sigma} & -\D 
		\end{pmatrix},
		\label{Dirac}
	\end{align}
	which we call the isotropic Wolff Hamiltonian.
	The wave vector $\bk$ is measured from $\bk_0$, where the bands have their extrema.
	(Such a situation is realized at the $L$-point in bismuth.)	
	The spin-orbit interaction is included in the off-diagonal elements and its magnitude is proportional to $\gamma$.
	The parabolic part $k^2/2m$ is negligibly small, when the spin-orbit interaction is much larger than the band gap (this is true for bismuth).

	Under an external magnetic field, the eigen energy can be easily obtained by the squared wave equation:
	\begin{align}
	\mathscr{H}^2 \psi =
	\left[
	\D^2 + 2\D
		\begin{pmatrix}
			\mathscr{H}^* & 0 \\
			0 & \mathscr{H}^*
		\end{pmatrix}
	\right]
		\psi = E^2 \psi.
	\end{align}
	Here the $2\times2$ Hamiltonian, $\mathscr{H}^*$,
	\begin{align}
		\mathscr{H}^* 
		&=\frac{\bpi^2}{2m_{\rm c}^*} + \frac{g^*\mu_{\rm B}}{2}\bm{\sigma}\cdot\bm{B}
		\label{effective Hamiltonian}
	\end{align}
	is the same as that for a free particle in a magnetic field ($\bpi = \bm{p}+e\bm{A}/c$) with the cyclotron mass $m_{\rm c}^*=\D/\gamma^2$ and the g-factor $g^*=2m/m_{\rm c}^*$.
	($\mu_{\rm B}=e/2mc$ is the Bohr magneton, and $e>0$.)
	The eigen energy of $\mathscr{H}^*$ is exactly given in the well-known manner, and then we have
	\begin{align}
	\pm E_{n,\sigma}(k_z)&= \pm\sqrt{\D^2 + 2\D \left[ \{n+(1+\sigma)/2\}\wc+ k_z^2 /2m_{\rm c}^*\right]},
	\label{ene} 
\end{align}
	where the magnetic field is along the $z$-direction, $\wc=eB/m_{\rm c}^* c$, and $\sigma=\pm 1$ is the eigenvalues of $\sigma_z$.
	The plus and minus sign in eq. (\ref{ene}) correspond to the conduction and valence band, respectively.
	The energy levels given by eq. (\ref{ene}) with $k_z=0$ are shown in Fig. \ref{LL} (a).

	It should be emphasized here that the conduction and valence band have different $B$-dependences in eq. (\ref{ene}).
	In the conduction band, the energy for $\sigma=+1$ (up spin) increases, while that for $\sigma=-1$ (down spin) decreases (see Fig. \ref{LL} (a)).
	In the valence band, on the other hand, the energy for $\sigma=+1$ decreases, while that for $\sigma=-1$ increases.
	In other words, the sign of the g-factor is opposite between the conduction and valence bands.
	Furthermore, the spin and cyclotron masses are the same, so that the spin splitting is equal to the orbital splitting and becomes very large since $m_{\rm c}^*$ is small.
	Then, the lowest Landau level is clearly separated from the opposite spin levels even by a very weak field.

	The magneto-optical conductivity tensor is given on the basis of the Kubo formula in the form\cite{Fukuyama1969b}
	\begin{align}
	\sigma_{\mu\nu} (\omega)
	&= \frac{1}{\Im \omega}
	\left[
	\Phi_{\mu\nu} (\omega + \Im \delta)-\Phi_{\mu \nu} (0+\Im \delta)
	\right],
	\\
	\Phi_{\mu \nu} (\Im \omega_\lambda)
		&=-e^2 T\sum_{n, i, j}
		\langle i | v_\mu | j\rangle
		\langle j | v_\nu | i\rangle
		\scr{G}_i (\Im \tilde{\ve}_n)
		\scr{G}_j (\Im \tilde{\ve}_n- \Im \omega_\lambda),
		\label{eq(5)}
\end{align}
	where $\bm{v}=\partial \mathscr{H}/\partial \bk$ is the velocity operator, $\scr{G}_i (\Im \ve_n)= \langle i |\left[ \Im \ve_n - \scr{H}\right]^{-1} | i \rangle$ is the Green function, and $|i \rangle$ is the state in which the energy is given by $E_i(k_z)$.
	The effect of damping is expressed by $\Gamma$ as $\tilde{\ve}_n=\ve_n + {\rm sgn}(\ve_n) \Gamma $.
	%
	For the inter-band (valence $\to$ conduction) transition, we have
\begin{align}
	&\Phi_{xx}^{\rm cv} =
	\frac{e^2\gamma^4 N_{\rm L}}{8}
	\sum_{n, k_z, \sigma}
	\Biggl[ 
	F(\omega, -E_{n, \sigma}, E_{n+1, \sigma})
	(A_{n+1 \sigma, n \sigma}^{\rc \rv })^2
	\nonumber\\&\times
	m \wc (n+1)
	\left[
	(E_{n+1, \sigma}-E_{n, \sigma} + 2\D ) + \sigma (E_{n+1, \sigma}+E_{n, \sigma})
	\right]^2 
	\nonumber\\
	&+
	F(\omega, -E_{n, \sigma}, E_{n-1, \sigma})
	(A_{n-1 \sigma, n \sigma}^{\rc \rv })^2
	\nonumber\\&\times
	m \wc n
	\left[
	(E_{n-1, \sigma}-E_{n, \sigma} + 2\D ) - \sigma (E_{n-1, \sigma}+E_{n, \sigma})
	\right]^2
	\nonumber\\
	&+2
	F(\omega, -E_{n, \sigma}, E_{n, -\sigma})
	( A_{n -\sigma, n \sigma}^{\rc \rv})^2 
	k_z^2 (E_{n, -\sigma}+E_{n, \sigma})^2
	\Biggr],
	%
	\label{Pxx_inter}
\end{align}
	where $N_{\rm L}=eB/2\pi c $ is the degrees of the Landau-level degeneracy, $A_{n \sigma, n' \sigma'}^{\rc \rv}= \left[ E_{n \sigma} E_{n' \sigma'} (E_{n \sigma} +\D) (E_{n' \sigma'}+\D)\right]^{-1/2}$, and $E_{n, \sigma}$ implicitly includes the $k_z$ dependences as $E_{n, \sigma}(k_z)$.
	The energy $-E_{n, \sigma}$ corresponds to the initial state in the valence band, and $E_{n\pm1, \sigma}$ and $E_{n, -\sigma}$ to the excited state in the conduction band.
	We performed the analytic continuation of the Green function part $\scr{G}(\Im \tilde{\ve}_n)\scr{G}(\Im \tilde{\ve}_n- \Im \omega_\lambda)$, and obtained the analytical form at $T=0$ as
	\begin{align}
		&F (\omega, E, E')
		=\frac{\Im}{2\pi}\Biggl[
	\frac{1}{\omega + E - E' + 2\Im \Gamma}
		\nonumber\\&\times
		\left(
		\ln \frac{\mu-E-\Im \Gamma}
		{\mu - \omega -E-\Im \Gamma}
		+
		\ln \frac{\mu-E'+\Im \Gamma}
		{\mu + \omega -E'+\Im \Gamma}
		\right)
		\nonumber\\
		&-\frac{1}{\omega+E-E'}
		\left(
		\ln \frac{\mu -E +\Im \Gamma}
		{\mu - \omega - E - \Im \Gamma}
		+
		\ln \frac{\mu -E' -\Im \Gamma}
		{\mu + \omega - E' + \Im \Gamma}
		\right)
		\Biggr].
\end{align}
	%
	%
	For the intra-band (conduction band) transition, we have the result just by replacing $-E_{n, \sigma} \to +E_{n, \sigma}$ in eq. (\ref{Pxx_inter}). 
	In eq. (\ref{Pxx_inter}),  each term expresses the transition $-E_{n\sigma} \to E_{n\pm1, \sigma}$ (orbital transition) and $-E_{n\sigma} \to E_{n, -\sigma}$ (spin transition), which are all $\D j = \pm 1$ transitions; 
	this is the selection rule for $\sigma_{xx}$.
	 (The quantum numbers $n$ and $\sigma$ are no more good quantum numbers, but only $j$ is the good quantum number.)
	%
\begin{figure}
\begin{center}
\includegraphics[width=80mm]{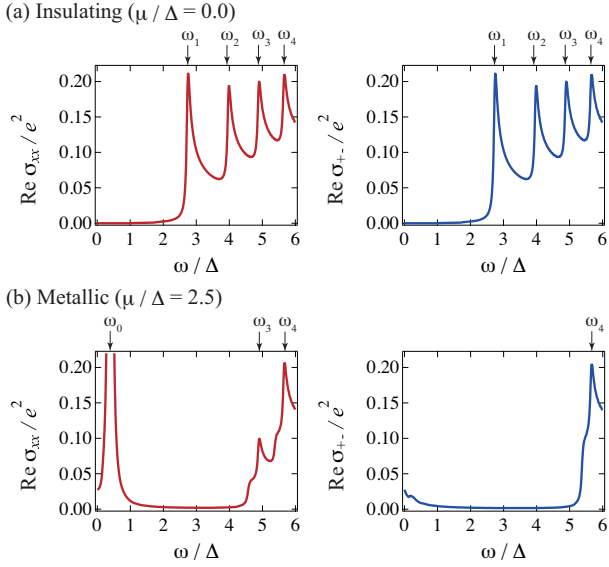}
\end{center}
\caption{(color online) Frequency dependence of Re $\sigma_{xx}$ (left) and Re $\sigma_{+-}$ (right) for (a) $\mu /\D = 2.5$ and (b) $\mu/\D = 0.0$.
}
\label{CxxCpm}
\end{figure}
	%
	
	The results of $\sigma_{xx}(\omega)$ are shown in the left of Fig. \ref{CxxCpm} for (a) insulating ($\mu=0.0$) and (b) metallic ($\mu/\D=2.5$) cases.
	All calculations are carried out for the whole processes ($\Phi^{\rc\rc} + \Phi^{\rv\rv} + \Phi^{\rc\rv} + \Phi^{\rv\rc}$) with $\wc / \D = 1.0$ and $\Gamma / \D = 0.02$.
	Each peak structure can be understood from the selection rule above.
	For the insulating region, $\mu/\D=0.0$, all peaks appear in the high energy ($\omega>2\D$) region, which are the contribution from the inter-band transitions of $(j=0_\rv \to 1_\rc)$ and $(j=1_\rv \to 0_\rc)$ at $\omega_1$, $(j=1_\rv \to 2_\rc)$ and $(j=2_\rv \to 1_\rc)$ at $\omega_2$, etc. (Fig. \ref{LL} (c)).
	(The subscripts denote the conduction (\rc) and valence (\rv) band levels, and $\omega_{j}=E_{j-1}(k_z =0) + E_{j}(k_z=0)$ as is indicated by arrows in Fig. \ref{CxxCpm}.)
	For the metallic region, $\mu/\D = 2.5$, the high energy peaks are due to the inter-band transitions of $(j=2_\rv \to 3_\rc)$ at $\omega_3$, $(j=3_\rv \to 4_\rc)$ and $(j=4_\rv \to 3_\rc)$ at $\omega_4$, etc. (Fig. \ref{LL} (d), $\mu$ locates between $j=2$ and 3 in the conduction band).
	The low energy ($\omega < \D$) sharp peak is the contribution from the intra-band transition of $(j=2_\rc \to 3_\rc)$. 
	In both regions, the excited states consist of both spins at every $\omega_j$, so that the current is not spin-polarized.
	Therefore, in $\sigma_{xx}(\omega)$, remarkable spin polarization cannot be obtained.
	However, the situation completely changes if we see the response to the circularly polarized light.

	The response to the circularly polarized light can be calculated by the same manner as in $\sigma_{xx}$ in terms of $v_\pm \equiv (v_x \pm \Im v_y)/\sqrt{2}$.
	The inter-band contribution to $\Phi_{+-}$ is then obtained as
\begin{align}
	&\Phi_{+-}^{\rc \rv} =
	\frac{e^2 \gamma^4N_{\rm L}}{4}
	\sum_{n, k_z, \sigma}\Biggl[
	F(\omega, -E_{n, \sigma}, E_{n-1, \sigma})
	(A_{n-1 \sigma, n \sigma}^{\rc \rv })^2
	\nonumber\\&\times
	m \wc n
	\left[
	(E_{n-1, \sigma}-E_{n, \sigma} + 2\D ) - \sigma (E_{n-1, \sigma}+E_{n, \sigma})
	\right]^2
	\nonumber\\
	&+
	2F(\omega, -E_{n, +1}, E_{n, -1})
	( A_{n, -1; n, +1 }^{\rc \rv})^2 
	k_z^2 (E_{n, +1}+E_{n, -1})^2
	\Biggr].
	\label{P+-}
\end{align}
	In this case, we have only the transitions of $\D j=-1$.
	The results of $\sigma_{+-}(\omega)$ are shown in the right of Fig. \ref{CxxCpm}.
	At first sight, there is no essential difference between $\sigma_{xx}(\omega)$ and $\sigma_{+-}(\omega)$, but their spin structures are definitely different.
	If we see the lowest excited state at $\omega_1$ in the insulating region, only  $(j=1_\rv \to 0_\rc)$ transition is possible for $\sigma_{+-}$, while both $(j=0_\rv \to 1_\rc)$ and $(j=1_\rv \to 0_\rc)$ are possible for $\sigma_{xx}$ (Fig. \ref{LL} (c)).
	All electrons excited by circularly polarized light of $\omega=\omega_1$ are concentrated at the lowest Landau level with a particular spin direction, i.e., the spin is polarized.
	Such a transition can be realized due to the spin transition in addition to the orbital transition.
	If we have orbital transition only, both spins are excited into the $j=0_\rc$ level by $\omega_1$, since the initial state $j=0_\rv$ consists of both spins, and their spins are conserved only by the orbital transition.
	In the present case, however, the spin-transition is also possible due to the spin-orbit coupling, so that the orbital ($1_\rv^- \to 0_\rc^-$) and spin ($0_\rv^+ \to 0_\rc^-$) transition lead the excited state consisting of down spin only.
	Therefore, we can obtain the spin-polarized current by the circularly polarized light.
	%
	%

\begin{figure}
\begin{center}
\includegraphics[width=80mm]{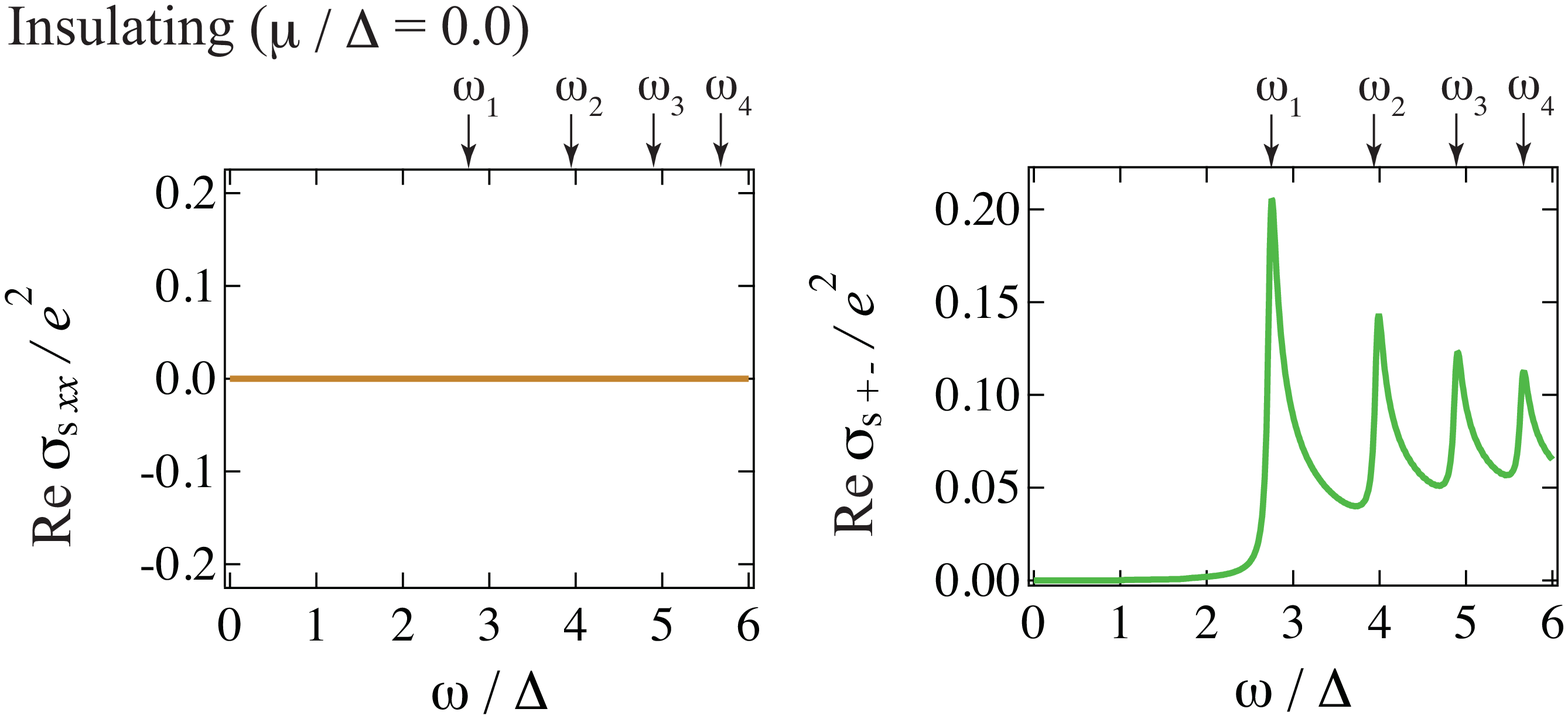}
\end{center}
\caption{(color online) Frequency dependence of Re $\sigma_{{\rm s}xx}$ (left) and Re $\sigma_{{\rm s}+-}$ (right) for $\mu /\D = 0.0$.}
\label{CsxxCspm}
\end{figure}
	%
	In order to verify the spin polarization of $\sigma_{+-}(\omega)$, let us calculate the current of the magnetic moment, which originates from the spin of excited electrons.
	As discussed before, the sign of g-factor in the conduction and valence band are opposite, so that we define the magnetic moment $\bm{\mu}_{\rm e}$ as
	\begin{align}
		\bm{\mu}_{\rm e}=-g^* \mu_{\rm B}
		\begin{pmatrix}
			\bm{\sigma} & 0 \\
			0 & -\bm{\sigma} 
		\end{pmatrix}.
		\label{magnetic moment}
	\end{align}
	In addition, we define the velocity of the magnetic moment along the $z$-direction by $v_{{\rm s} i}=\mu_{{\rm e}z} v_i/g^* \mu_{\rm B}$ ($i=x, y$), which is an Hermitian operator.
	(Here, $v_i$ being the velocity operator introduced in eq. (\ref{eq(5)}))
	Then, the spin-polarization of the current can be estimated with this velocity of the magnetic moment as $\Phi_{{\rm s}\mu \nu}=-e^2 T\sum_{n, i, j} \langle i | v_{{\rm s}\mu} | j \rangle\langle j | v_{\nu} | i \rangle \scr{G}_i(\Im \tilde{\ve}_n) \scr{G}_j(\Im \tilde{\ve}_n- \Im \omega_\lambda)$.
	It should be noted that the spin current is not well defined since the spin density is not a conserved quantity in the presence of spin-orbit coupling.
	Here we use $\Phi_{{\rm s}\mu \nu}$ just to see the tendency of the spin-polarization.
	The responses $\Phi_{{\rm s}xx, {\rm s}+-}$ are calculated in the same manner as $\Phi_{xx, +-}$, and their results are shown in Fig. \ref{CsxxCspm} for the insulating region.
	%

	%
	%
	The spin polarization is evaluated by $|\sigma_{{\rm s}\mu\nu}|/\sigma_{\mu\nu}$.
	Therefore, Figs. \ref{CxxCpm} and \ref{CsxxCspm} signify that $\sigma_{+-}(\omega)$ in the insulating region is highly spin-polarized, while $\sigma_{xx}(\omega)$ is not.
	Especially, at the lowest excitation energy, $|\sigma_{{\rm s}+-}(\omega_1)|/\sigma_{+-}(\omega_1) = 1$, namely, we have 100\% spin polarization as is expected. 
	(More precisely, the spin-polarization becomes slightly less than 100\% for finite $\Gamma$, e.g., $|\sigma_{{\rm s}+-}|/\sigma_{+-}=0.97$ at $\omega_1$ for $\Gamma /\D= 0.02$.)
	%
	%
	%
	%

	%
	One might think that the spin relaxation is very fast in cases with a strong spin-orbit coupling, thereby destroying the spin-polarization, but this is not the case here.
	The Landau levels under consideration are the pure eigenstates of the Wolff Hamiltonian, which already includes the effect of strong spin-orbit coupling, so that different Landau levels are not mixed with each other, i.e., the quantum number $j$ conserves.
	Moreover, the present spin-polarized current uses the lowest Landau level, whose spin is unique and is not mixed, so that the spin-polarization is kept.

	Now we discuss the implications of the present results on the experimental measurements.
	Our spin-polarized current flows through the bulk not the surface current as in the topological insulators.\cite{HZhang2009} 
	The magnitude of the current is large, and it should be much easier to detect.
	Actually, the magneto-optical measurements on bismuth already exhibit clear peak structures\cite{Vecchi1976}.
	Therefore, the present proposal of spin-polarized current will be confirmed experimentally, e.g., by the {\it ordinary} magneto-optical measurements with the use of the Kerr rotation microscopy\cite{Kato2007}.
	There, we will see the bulk spin polarization through the photo-induced change in magnetization.
	%
	
	%
	The spin-orbit interaction of bismuth ($\sim 1.5$ eV\cite{Liu1995}) is the largest among the non-radioactive elements, and much larger than the band gap $2\D \simeq 15$ meV.
	This is the ideal situation of the Wolff Hamiltonian.
	The cyclotron mass is extremely small $m_\rc^* \simeq 0.002$ ($B\parallel$ bisectrix axes\cite{Zhu2011}), and the g-factor exceeds 1000, which enables the huge spin splittings.
	The situation in Fig. \ref{CxxCpm} ($\wc = \D$) is achieved only by $B=0.12$ T. 
	The insulating Dirac electrons appear in Bi$_{1-x}$Sb$_x$ alloys ($0.07<x<0.22$)\cite{Wehrli1968,Brandt1977,Lenoir1996}. 
	Also, Ca$_3$PbO and its family would be good candidates for the insulating Dirac electron\cite{Kariyado2011}.

	We discussed the transport phenomena with the isotropic Wolff Hamiltonian in this Letter.
	However, the velocity of bismuth is highly anisotropic, so that it would be important to take into account the anisotropy when we compare with experimental results\cite{Vecchi1976,Zhu2011}.
	Generally speaking, if we consider the anisotropy, other matrix elements can appear, which will give additional orbital- and spin-transition terms.
	The clarification of these consequences needs further studies.
	%
	%

	%
	The Wolff Hamiltonian is essentially different from the graphene Hamiltonian with respect to the spin properties.
	In the Wolff Hamiltonian, the signs of g-factor for the real spin of electrons are opposite between the conduction and valence band, while they are the same in the graphene Hamiltonian.
	Furthermore, the spin splitting is extremely large and the lowest Landau level includes only a single spin with a particular direction in the former, while the spin splitting is tiny and the lowest Landau level is spin degenerate in the latter.
	Therefore, the spin-polarized current discussed in this Letter is only possible in the Wolff Hamiltonian, while it is difficult to realize in the graphene Hamiltonian.
	Nevertheless, the spin properties of the Wolff Hamiltonian are closely related to the pseudo-spin properties of the graphene Hamiltonian\cite{Koshino2010,Koshino2011}, so that an interesting pseudo-spin-polarized current may be realized by the present mechanism in the graphene Hamiltonian.

	We have predicted the spin-polarized electric current in the isotropic Wolff model by calculating $\sigma_{xx}(\omega)$ and $\sigma_{+-}(\omega)$ on the basis of the Kubo formula.
	It has been shown that the spin-polarized current is possible by the circularly polarized light in the insulating regions. 
	The keys to achieving the spin polarization are 
	(i) the well-isolated lowest Landau level originated from the large spin splitting;
	(ii) the non-trivial spin transition in addition to the usual orbital transition; and
	(iii) the opposite sign of the g-factor or the magnetic moment between the conduction and valence band.
	These points are originated from the strong spin-orbit interaction, and are the prominent characteristics of the Dirac electrons with spin-orbit interactions.
	%
	%
	It would be exceedingly interesting to test these ideas by magneto-optical measurements with the use of Kerr rotation microscopy.

	We thank J. Inoue, K. Miyake and H. Kohno for helpful comments.
	This work is financially supported by MRL System, and by ``HISHO" The Top Thirty Young Researchers of Osaka University Project, and also by Young Scientists (B) (No. 23740269) from Japan Society for the Promotion of Science.
	%

\bibliographystyle{jpsj}
\bibliography{Bismuth,footnote}
\end{document}